\documentclass[preprint,eqsecnum,aps,nofootinbib]{revtex4}
\usepackage{amsfonts,amsmath,amssymb,amsthm}
\usepackage{latexsym}
\usepackage{bbm,bm}
\usepackage{graphicx}

\newcommand{\ket}[1]{\lvert #1 \rangle}
\newcommand{\bra}[1]{\langle #1 \lvert}
\newcommand{\beq}{\begin{equation}}
\newcommand{\eeq}{\end{equation}}
\newcommand{\beqs}{\begin{eqnarray}}
\newcommand{\eeqs}{\end{eqnarray}}
\begin{document}

\title{Feynman Propagator of the Arthurs-Kelly system at the Planck Scale}

 \author{ Mi-Ra Hwang$^{1,2}$, Eylee Jung$^{1,2}1$, MuSeong Kim$^3$ and DaeKil Park$^{1,2,4}$\footnote{corresponding author, dkpark@kyungnam.ac.kr} }

\affiliation{$^1$ Quanta, Changdong 76, Masan, 51730, Korea     \\
                $^2$Department of Electronic Engineering, Kyungnam University, Changwon,
                 631-701, Korea    \\
                 $^3$Pharos iBio Co., Ltd. 
                Head Office: \#1408, 38, Heungan-daero 427beon-gil, Dongan-gu, Anyang, 14059, Korea \\
                $^4$Department of Physics, Kyungnam University, Changwon,
                  631-701, Korea }

\preprint{\bf{KMC-25-01}}

\begin{abstract}
The non-relativistic quantum mechanics with a generalized uncertainty principle (GUP) is examined in the Arthurs-Kelly system. 
The Feynman propagator for this system is exactly derived within the first order of the GUP parameter $\beta$. 
The application of it in the early universe stage is briefly discussed.

\end{abstract}

\maketitle

\section{Introduction}
Physics at the Planck scale suggests the existence of the minimal length (ML).  
The existence of the ML at this scale seems to be a universal characteristic of quantum gravity\cite{townsend76,amati89,garay94}. It appears in loop quantum gravity\cite{rovelli88,rovelli90,rovelli98,carlip01},
string theory\cite{konishi90,kato90,strominger91}, path-integral quantum gravity\cite{padmanabhan85,padmanabhan85-2,padmanabhan86,padmanabhan87,greensite91}, and black hole physics\cite{maggiore93}. 
ML also appeared in some microscope thought-experiment\cite{mead64}. 
From an aspect of quantum mechanics the existence of ML modifies the uncertainty principle from Heisenberg uncertainty principle (HUP)\cite{uncertainty,robertson1929}\footnote{Of course, in Eq. (\ref{HUP-1}) $\left( \Delta_{\hat{A}}^2 \right)_{\psi} \equiv \bra{\psi} \hat{A}^2 \ket{\psi} - \bra{\psi} \hat{A} \ket{\psi}^2$ is a variance computed in the measurement of the observable $\hat{A}$ when the state vector is $\ket{\psi}$.
We will drop the quantum state $\ket{\psi}$ when it is evident in a context.}
\begin{equation}
\label{HUP-1}
\Delta_{\hat{x}} \Delta_{\hat{p}} \geq \frac{\hbar}{2}
\end{equation}
to an generalized uncertainty principle (GUP). This is because of the fact that the uncertainty of the position should be larger than the ML. 
Although there are several expressions on the GUP, we use in this paper the simplest form of the GUP proposed in Ref.\cite{kempf93,kempf94}
\begin{equation}
\label{GUP-d-1}
\Delta_{\widehat{P}_i} \Delta_{\widehat{Q}_i}  \geq \frac{\hbar}{2} \left[ 1 + \beta \left(\Delta_{\widehat{{\bf P}}^2} + \langle \widehat{{\bf P}} \rangle^2 \right) + 2 \beta \left( \Delta_{\widehat{P}_i^2} + \langle \widehat{P}_i \rangle^2 \right) \right]
\hspace{1.0cm}  (i = 1, 2, \cdots, D)
\end{equation}
or equivalently
\begin{eqnarray}
\label{GUP-d-2}
&& \left[ \widehat{Q}_i, \widehat{P}_j \right] = i \hbar \left( \delta_{ij} + \beta \delta_{ij} \widehat{{\bf P}}^2 + 2 \beta \widehat{P}_i \widehat{P}_j  \right)    \\    \nonumber
&& \hspace{1.0cm} \left[ \widehat{Q}_i, \widehat{Q}_j \right] = \left[\widehat{P}_i, \widehat{P}_j \right] = 0
\end{eqnarray}
where $\beta$ is a GUP parameter, which has a dimension $(\mbox{momentum})^{-2}$. 
The existence of the ML can be pictorially shown at $D=1$ as shown in Ref.\cite{park20-1}.

If $\beta$ is small, Eq. (\ref{GUP-d-2}) can be solved up to ${\cal O} (\beta)$ as 
\begin{equation}
\label{GUP-d-4}
\widehat{P}_i = \widehat{p}_i \left[1 + \beta \widehat{{\bf p}}^2  + {\cal O} (\beta^2) \right]   \hspace{1.0cm} \widehat{Q}_i = \widehat{q}_i
\end{equation}
where $\widehat{p}_i$ and $\widehat{q}_i$ obey the usual HUP. Using Eq. (\ref{GUP-d-4}) and Feynman's path-integral technique\cite{feynman,kleinert} the Feynman propagator (or kernel) was exactly derived 
up to ${\cal O} (\beta)$ for $D=1$ free particle case\cite{das2012,gangop2019}. Also the propagator for simple harmonic oscillator (SHO) was also derived in Ref. \cite{comment-1,park20-1}. Furthermore, the peculiar 
property of the Feynman propagator was discussed in Ref. \cite{park2020} in the GUP-corrected quantum mechanics when the potential is singular such as $\delta$-function potential.
The effect of the ML is discussed in Aharonov-Bohm scattering\cite{p135}, R\'{e}nyi and von Neumann entropies of thermal state \cite{p136}, and quantum entanglement\cite{p138}.

In this paper we will derive the Feynman propagator for Arthurs-Kelly (AK) system within first order of $\beta$ in the GUP-corrected quantum mechanics.  
In Sec. II we will review the AK system briefly. In this section we will discuss why the explicit derivation of the Feynman propagator is difficult in the GUP-corrected AK system.
In Sec. III factorization of the time-evolution for the AK system is carried out by making use of the Baker-Campbell-Hausdorff (BCH) formula. 
In Sec. IV the Feynman propagator for the GUP-corrected AK system is explicitly derived.
In Sec. V a brief conclusion is given. 
In appendix A main calculation for the factorization of the time-evolution operator is explicitly presented.
In appendix B we summarize the coefficients arising in the Feynman propagator.

\section{Brief Review of the Arthurs-Kelly System}
For a long time\footnote{$\hbar=1$ will be used in the following for simplicity.} it is believed that Eq. (\ref{HUP-1}) states that the complementary variables such as position and momentum cannot be simultaneously measured with arbitrary precision\cite{cohen}. 

However, recent study indicates that the inequality (\ref{HUP-1}) is not related to the simultaneous measurement.
This can be understood from the fact that Eq. (\ref{HUP-1}) does not involve the effect of the measurement apparatus. Rather, the interpretation of Eq. (\ref{HUP-1}) is related to a restriction on quantum state preparation\cite{aguilar18,ballentine70,Buss10,ozawa89,appleby98,raymer94}.
This is summarized in Ref.\cite{text}, where the following comment is presented: {\it The correct interpretation of the UP is that if we prepare a large number of quantum systems in identical state, $\ket{\psi}$, and then perform measurements of 
$\hat{x}$ on some of those systems, and of $\hat{p}$ in others, then the standard deviation $\Delta_{\hat{x}}$ times the standard deviation $\Delta_{\hat{p}}$ will satisfy the inequality (\ref{HUP-1})}.

The first quantum-mechanical description of the joint measurement of complementary variables came up with a paper by AK \cite{Ar-Kelly}. 
They generalized the von Neumann measurement process\cite{neumann32}, where a position measurement was considered via a single additional pointer system. For the simultaneous measurement  AK introduced two probes for the measurement process. Thus, the whole system consists of three parties: one of them which is ``3'' in the following is 
a physical particle, whose position and momentum will be jointly measured, and two of them which are ``1'' and ``2'' are the pointer systems. The interaction Hamiltonian is chosen as a form
${\widehat H}_{int} = \kappa \left({\hat x}_3 {\hat p}_1 + {\hat p}_3 {\hat p}_2 \right)$ with large coupling constant $\kappa$.  It is assumed that the joint measurement is performed at time $t = 1 / \kappa$ for simplicity.
Then, AK showed that if the initial state is written as $\Psi_{in} (x_1, x_2, x_3: t=0) = \phi_1(x_1) \phi_2 (x_2) \phi_3 (x_3)$ and each $\phi_j$ is a Gaussian state with minimum uncertainty,
the uncertainty for the meters holds the inequality 
\begin{equation}
\label{Ar-K-1}
\Delta_{\hat{x}_1} (t = 1 / \kappa) \Delta_{\hat{x}_2} (t = 1 / \kappa) \geq 1.
\end{equation}
This is twice of the Heisenberg uncertainty given in Eq. (\ref{HUP-1}). The experimental test of the joint measurement was also discussed in Ref. \cite{exp1,exp2,exp3}. 

Recently, defining a total Hamiltonian
\begin{equation}
\label{hamil-1}
\widehat{H} = \widehat{H}_{free} + \widehat{H}_{int}
\end{equation}
where
\begin{equation}
\label{hamil-2}
\widehat{H}_{free} = \frac{\hat{p}_1^2}{2 m_1} + \frac{\hat{p}_2^2}{2 m_2} +  \frac{\hat{p}_3^2}{2 m_3} 
\hspace{1.0cm} \widehat{H}_{int} = \kappa \left(\hat{x}_3 \hat{p}_1 + \hat{p}_3 \hat{p}_2 \right),
\end{equation}
the Feynman propagator of this system is derived in Ref. \cite{p143} as a following form:
\begin{equation}
\label{feynman1} 
K_{AK}[ Q_1, Q_2, Q_3: q_1, q_2, q_3: t] =  \sqrt{\frac{3 m_1 m_2 m_3^2}{2 \pi^3 i b t^3 a(t)}} e^{i S_{cl}}
\end{equation}
where 
\begin{eqnarray}
\label{cl_action1}
&& S_{cl} = \frac{1}{2 a(t) b t} \Bigg[ 12 m_1 m_3 b z_1^2 - m_2 a(t) z_2^2 - m_3 a(t) z_-^2 + 3 m_1 m_3 \kappa^2 t^2 b z_+^2         \\   \nonumber
&&   \hspace{4.0cm}   - 12 m_1 m_3 \kappa t b z_1 z_+ + 2 m_2 m_3 \kappa a(t) z_2 z_- \Bigg].
\end{eqnarray}
In Eq. (\ref{feynman1}) and Eq. (\ref{cl_action1}) $b = m_2 m_3 \kappa^2 - 1$,  $a(t) = 12 m_3 + m_1 \kappa^2 t^2$ and 
\begin{equation}
\label{eqofmotion3}
z_1 = Q_1 - q_1    \hspace{.5cm}  z_2 = Q_2 - q_2  \hspace{.5cm} z_{\pm} = Q_3 \pm q_3 .
\end{equation} 
Using the Feynman propagator the effect of the initial entanglement in the pointer system is examined in the AK inequality in Ref. \cite{p143}.

Before we derive the Feynman propagator or Kernel in the GUP-corrected AK system, let us briefly comment how Eq. (\ref{feynman1}) is derived. 
Usually the Kernel can be derived by direct path integration
\begin{equation}
\label{path-def}
K[q_f, t_f: q_0, t_0] \equiv \bra{q_f, t_f} q_0, t_0 \rangle = \int_{(t_0, q_0)}^{(t_f, q_f)} {\cal D} q e^{i  S[q]},
\end{equation}
where $S[q]$ is an actional functional and ${\cal D} q$ is sum over all possible paths connecting $(t_0, q_0)$ and $(t_f, q_f)$ in spacetime. 
If the direct path-integration is extremely difficult, we can adopt another method by using the relation between Kernel and time-independent Schr\"{o}dinger equation as follows;
\begin{equation}
\label{path-def-2}
K[q_f, t_f: q_0, t_0] = \sum_{n=1}^{\infty} \phi_n (q_f) \phi_n^* (q_0) e^{-i E_n (t_f - t_0)},
\end{equation}
where $\phi_n (q)$ and $E_n$ are eigenfunction and eigenvalue of Schr\"{o}dinger equation. If we use the latter, we have to solve the following Schr\"{o}dinger equation;
\begin{eqnarray}
\label{schrodinger1}
&& \left[-\frac{1}{2 m_1} \frac{\partial^2}{\partial q_1^2} -\frac{1}{2 m_2} \frac{\partial^2}{\partial q_2^2} -\frac{1}{2 m_3} \frac{\partial^2}{\partial q_3^2} - \kappa \left(i q_3 \frac{\partial}{\partial q_1} +   \frac{\partial}{\partial q_2} \frac{\partial}{\partial q_2} \right)  \right]
\Phi (q_1, q_2, q_3)                                                                             \\   \nonumber
&&  \hspace{9.0cm}   = E(p_1, p_2, p_3) \Phi(q_1, q_2, q_3).
\end{eqnarray}
However, it seems to be highly difficult, at least to us,  to derive the eigenfunction $\Phi(q_1, q_2, q_3)$ and eigenvalue $E(p_1, p_2, p_3)$ from Eq. (\ref{schrodinger1}).
Fortunately, the corresponding action for the AK system is quadratic.Thus we used the property of the direct path-integration, which states that when action is quadratic, the Kernel is represented by prefactor $F(t)$ and classical action $S_{cl}$.

Now, let us consider the GUP-corrected AK system. For the computation of the Feynman propagator up to ${\cal O} (\beta)$ we have to use Eq. (\ref{GUP-d-4}). 
Then, the corresponding action is not quadratic any more. Furthermore, the corresponding time-independent Schr\"{o}dinger equation is more complicated than Eq. (\ref{schrodinger1}). 
Thus, it seems to be highly difficult to derive the Feynman propagator exactly for the GUP-corrected AK system by using direct path-integration or  time-independent Schr\"{o}dinger equation.
Hence, in this paper we use a different method, where the time-evolution operator is used. Since the Feynman propagator is a coordinate representation of the time-evolution operator, we need to properly factorize the operator.
The factorization scheme is discussed in next section.

\section{Time-evolution operator of the AK System}

Let us consider the Hamiltonian in the GUP-corrected AK system
\begin{equation}
\label{hamil-1}
\widehat{H} = \frac{\widehat{P}_1^2}{2 m_1} + \frac{\widehat{P}_2^2}{2 m_2} +  \frac{\widehat{P}_3^2}{2 m_3} + \kappa \left(\widehat{Q}_3 \widehat{P}_1 + \widehat{P}_3 \widehat{P}_2 \right)
\end{equation}
where $\widehat{P_i}$ and $\widehat{Q_i}$ obey the one-dimensional version of Eq. (\ref{GUP-d-2}).
Applying Eq. (\ref{GUP-d-4}) the Hamiltonian can be represented by $\{q_i, p_i \}$ up to ${\cal O}(\beta)$ as following expression:
\begin{eqnarray}
\label{hamil-2}
&&\widehat{H} =  \frac{\widehat{p}_1^2}{2 m_1} + \frac{\widehat{p}_2^2}{2 m_2} +  \frac{\widehat{p}_3^2}{2 m_3} + \beta \left( \frac{\widehat{p}_1^4}{m_1} + \frac {\widehat{p}_2^4}{m_2} + \frac{ \widehat{p}_3^4}{m_3} \right)      \\   \nonumber
&&\hspace{.5cm}  + \kappa \bigg[ \left( \widehat{q}_3 \widehat{p}_1 + \widehat{p}_3 \widehat{p}_2 \right) + \beta \left( \widehat{q}_3 \widehat{p}_1^3 + \widehat{p}_2^3 \widehat{p}_3 + \widehat{p}_2 \widehat{p}_3^3 \right) \bigg] + {\cal O}(\beta^2),
\end{eqnarray}
where $\widehat{p}_i$ and $\widehat{q_i}$ obey the usual HUP $[\widehat{q}_i, \widehat{p}_j]= i \delta_{ij}$.
In Appendix A we successfully factorize the time-evolution operator $U(T) = e^{-i \widehat{H} T}$ by applying the Baker-Campbell-Hausdorff (BCH) formula several times. 
The final expression is 
\begin{equation}
\label{unitary-1}
\widehat{U} (T) = e^{\Delta x_1 \widehat{p}_1^2 + \beta \Delta y_1 \widehat{p}_1^4} e^{-\frac{i T}{2 m_2} \widehat{p}_2^2 - \frac{i \beta T}{m_2} \widehat{p}_2^4} \widehat{V} (T) e^{-i \kappa T \widehat{q}_3 (\widehat{p}_1 + \beta \widehat{p}_1^3)} 
e^{-\frac{i \beta \kappa^4 T^5}{6 m_3} \widehat{p}_1^4} \widehat{V} (T)
\end{equation}
where 
\begin{eqnarray}
\label{unitary-2}
&&\Delta x_1 = -\frac{i T}{2 m_1} + \frac{i \kappa^2 T^3}{12 m_3}   \hspace{1.0cm} \Delta y_1 = - \frac{i T}{m_1} + \frac{i \kappa^2 T^3}{6 m_3} - \frac{i \kappa^4 T^5}{30 m_3}       \\     \nonumber
&&\widehat{V} (T) = e^{-\frac{i T}{4 m_3} \widehat{p}_3^2 - \frac{i \beta T}{2 m_3} \widehat{p}_3^4} e^{-\frac{i \kappa T}{2} \left[ \widehat{p}_2 \widehat{p}_3 + \beta (\widehat{p}_2^3 \widehat{p}_3 + \widehat{p}_2 \widehat{p}_3^3 ) \right]}
e^{\frac{i \beta \kappa^2 T^3}{2 m_3} \widehat{p}_1^2 \widehat{p}_3^2} e^{\frac{i \beta \kappa^3 T^3}{4} \widehat{p}_1^2 \widehat{p}_2 \widehat{p}_3}.
\end{eqnarray}
When $\beta = 0$, Eq. (\ref{unitary-1}) becomes
\begin{equation}
\label{unitary-3}
\widehat{U}_{AK} (T) = e^{\Delta x_1 \widehat{p}_1^2} e^{-\frac{i T}{2 m_2} \widehat{p}_2^2} e^{-\frac{i T}{4 m_3} \widehat{p}_3^2} e^{-\frac{i \kappa T}{2} \widehat{p}_2 \widehat{p}_3} e^{-i \kappa T \widehat{q}_3 \widehat{p}_1}
e^{-\frac{i T}{4 m_3} \widehat{p}_3^2} e^{-\frac{i \kappa T}{2} \widehat{p}_2 \widehat{p}_3}.
\end{equation}
This is exactly the same with Eq. (2.10) of Ref.\cite{free_evolution}, which is the time-evolution operator of the usual AK system. If $\kappa = 0$, Eq. (\ref{unitary-1}) reduces to 
\begin{equation}
\label{unitary-4}
\widehat{U}_F (T) =  e^{-\frac{i T}{2 m_1} \widehat{p}_1^2 - \frac{i \beta T}{m_1} \widehat{p}_1^4} e^{-\frac{i T}{2 m_2} \widehat{p}_2^2 - \frac{i \beta T}{m_2} \widehat{p}_2^4} e^{-\frac{i T}{2 m_3} \widehat{p}_3^2 - \frac{i \beta T}{m_3} \widehat{p}_3^4}.
\end{equation}
This is just the time-evolution operator for the three one-dimensional free particles in the GUP-corrected quantum mechanics.

\section{Feynman Propagator}
In this section we derive the Feynman propagator for the Hamiltonian (\ref{hamil-2}) up to ${\cal O}(\beta)$. Since the Feynman propagator is a coordinate representation of the time-evolution operator, we should compute $\bra{Q_1, Q_2, Q_2} \widehat{U} (T) \ket{q_1, q_2, q_3}$.
Using $\bra{q} p \rangle = \frac{1}{\sqrt{2 \pi}} e^{i q p}$, where $\widehat{q} \ket{q} = q \ket{q}$ and $\widehat{p} \ket{p} = p \ket{p}$, one can exchange $\ket{q} \leftrightarrow \ket{p}$ freely. Thus the Feynman propagator $\bra{Q_1, Q_2, Q_2} \widehat{U} (T) \ket{q_1, q_2, q_3}$
can be explicitly computed in principle. Since we need to use the complete condition five times, the Feynman propagator is represented by a quintuple integral. Among them, however, double integration can be analytically performed using a $\delta$-function, 
and as a result, it is represented as a following triple integral:
\begin{eqnarray}
\label{feynman-11}
&& \bra{Q_1, Q_2, Q_3} \widehat{U} (T) \ket{q_1, q_2, q_3}   \\    \nonumber
&&= \frac{1}{(2 \pi)^3} \int dp_1 dp_2 dp_3 \exp\bigg[ -A p_1^2 - B p_2^2 - C p_3^2 + 2 D_1 p_1 p_2 + 2 D_2 p_1 p_3             \\   \nonumber
&&  \hspace{7.0cm}     + 2 D_3 p_2 p_3 + F_1 p_1 + F_2 p_2 + F_3 p_3 \bigg]                                                                         \\    \nonumber
&&    \times  \Bigg[ 1 + \beta \bigg\{ G_1 p_1^4 - \frac{i T }{m_2} p_2^4 - \frac{i T}{m_3} p_3^4 - i \kappa T Q_3 p_1^3 + G_2 p_1^3 p_2 + G_3 p_1^3 p_3     \\   \nonumber
&&\hspace{1.5cm} - \frac{2 i \kappa^2 T^3}{m_3} p_1^2 p_3^2 - i \kappa^3 T^3 p_1^2 p_2 p_3 + \frac{2 i \kappa T^2}{m_3} p_1 p_3^3 + \frac{i \kappa^2 T^2}{2} p_1 p_2^3     \\   \nonumber
&& \hspace{2.0cm} + \frac{3 i \kappa^2 T^2}{2} p_1 p_2 p_3^2 - i \kappa T \left(p_2^3 p_3 + p_2 p_3^3 \right)         \bigg\}   + O(\beta^2)        \Bigg]
\end{eqnarray}
where 
\begin{eqnarray}
\label{feynman-12}
&&G_1 = -\frac{i T}{m_1} - \frac{i \kappa^2 T^3}{3 m_3} - \frac{i \kappa^4 T^5}{5 m_3}   \hspace{.5cm} G_2 = \frac{i \kappa^2 T^2}{2} + \frac{i \kappa^4 T^4}{4}   \hspace{.5cm} G_3 = \frac{i \kappa T^2}{2 m_3} + \frac{i \kappa^3 T^4}{m_3}   \\   \nonumber
&& A = \frac{i \kappa^2 T^3}{4 m_3} - \Delta x_1   \hspace{1.0cm} B = \frac{i T}{2 m_2}   \hspace{1.0cm}   C = \frac{i T}{2 m_3}                                           \\    \nonumber
&& D_1 = \frac{i \kappa^2 T^2}{4}     \hspace{1.0cm}   D_2 = \frac{i \kappa T^2}{4 m_3}      \hspace{1.0cm}  D_3 = - \frac{i \kappa T}{2}                               \\    \nonumber
&& F_1 = i (z_1 - \kappa t Q_3)     \hspace{1.0cm}   F_2 = i z_2     \hspace{1.0cm}   F_3 = i z_3
\end{eqnarray}
with $z_i \equiv Q_i - q_i$.
Using an integral formula
\begin{eqnarray}
\label{int_formula1}
&&\int dx_1 dx_2 dx_3 \exp\bigg[ -A x_1^2 - B x_2^2 - C x_3^2 + 2 D_1 x_1 x_2 + 2 D_2 x_1 x_3             \\   \nonumber
&&  \hspace{4.0cm}     + 2 D_3 x_2 x_3 + F_1 x_1 + F_2 x_2 + F_3 x_3 \bigg] = \frac{\pi^{3/2}}{\sqrt{\tilde{\alpha}}} e^{\tilde{\beta} / (4 \tilde{\alpha}) }
\end{eqnarray}
where 
\begin{eqnarray}
\label{int_formula2}
&& \tilde{\alpha} = ABC - 2 D_1 D_2 D_3 - A D_3^2 - B D_2^2 - C D_1^2                             \\   \nonumber
&& \tilde{\beta} = (BC - D_3^2 ) F_1^2 + (AC - D_2^2 ) F_2^2 + (AB - D_1^2 ) F_3^2 + 2 F_1 F_2 (C D_1 + D_2 D_3 )      \\   \nonumber
&& \hspace{2.0cm}   + 2 F_1 F_3 (B D_2 + D_1 D_3 ) + 2 F_2 F_3 (A D_3 + D_1 D_2 ),
\end{eqnarray}
the triple integral in Eq. (\ref{feynman-11}) can be explicitly carried out and the final expression is 
\begin{eqnarray}
\label{feynman-13}
&& \bra{Q_1, Q_2, Q_3} \widehat{U} (T) \ket{q_1, q_2, q_3}   \\    \nonumber                \\   \nonumber
&& = K_{AK} [\bm{Q}; \bm{q}; T] \left[ 1 + \frac{\beta}{5 T^3 b^3 a^4(T)}    \sum^4_{\begin{subarray}{c}i,j,k = 0 \\ i+j+k \leq 4   \end{subarray}}  f_{ijk} z_1^i z_2^j z_3^k  + {\cal O}(\beta^2) \right].
\end{eqnarray}
The following six coefficients are exactly zero; $f_{310} = f_{130} = f_{030} = f_{210} = f_{110} = f_{010} = 0$.
Some coefficients are too lengthy to express them explicitly in this paper. For example $f_{000} = -3 T^2 b^2 \left[u_1(T) Q_3^4 + u_2(T) Q_3^2 + u_3(T) \right]$, where
\begin{eqnarray}
\label{coeff-1}
&&u_1(T) = 432 i b m_1^4 m_3^3 \kappa^8 T^6                               \\    \nonumber
&&u_2 (T) = 12 m_1^2 m_3^2 \kappa^2 T a(T) \bigg[ 60 b \left\{ a(T) + m_3 \kappa^2 T^2 \right\}                     \\   \nonumber
&& \hspace{5.0cm} + \kappa^2 T^2 \left\{ 28 b m_1 \kappa^2 T^2 - 5 m_3 (12 + m_1 m_2 \kappa^4 T^2 ) \right\} \bigg]     \\   \nonumber
&&u_3 (T) = i a^2 (T) \Bigg[ 60 m_3 a(T) (m_2 + m_3 - m_1 b) - 60 m_1 m_3^2 (b - 2) \kappa^2 T^2                              \\    \nonumber
&&  \hspace{5.0cm}        + 5 m_1 m_2 \kappa^2 T^2 a(T) - 2 m_1^2 m_3 (7 b - 5) \kappa^4 T^4          \Bigg].
\end{eqnarray}
Some other coefficients which does not have long expressions are summarized in Appendix B. Also using Eq. (\ref{app-b-2}) one can show that when $\kappa = 0$, the Feynman propagator becomes
\begin{eqnarray}
\label{feynman_free}
&&\bra{Q_1, Q_2, Q_3} \widehat{U}_{F} (T) \ket{q_1, q_2, q_3}      \\    \nonumber
&& = \sqrt{\frac{i m_1 m_2 m_3}{8 \pi^3 T^3}} \exp \left[ i \frac{m_1 z_1^2 + m_2 z_2^2 + m_3 z_3^2}{2 T} \right] \Bigg[ 1 + \beta \sum_{j = 1}^3 \left\{ \frac{3 i m_j}{T} - \frac{6 m_j^2 z_j^2}{T^2} - \frac{i m_j^3 z_j^4}{T^3} \right\} + {\cal O}(\beta^2) \Bigg]  \\   \nonumber
&& =  \sqrt{\frac{i m_1 m_2 m_3}{8 \pi^3 T^3}} \exp \left[ \frac{i}{2 T} \sum_{j=1}^3 m_j z_j^2 \left\{ 1 - 2 \beta \left( \frac{m_j z_j}{T} \right)^2 \right\} \right]
\Bigg[ 1 + \beta \sum_{j = 1}^3 \left\{ \frac{3 i m_j}{T} - \frac{6 m_j^2 z_j^2}{T^2} \right\} + {\cal O}(\beta^2) \Bigg].
\end{eqnarray}
This is Feynman propagator for the three free particles in the GUP-corrected quantum mechanics. This is consistent with Ref. \cite{das2012,gangop2019,park2020}, where Feynman propagator for the single free particle is derived in the GUP-corrected quantum mechanics.

\section{conclusion}

We derive the Feynman propagator explicitly up to ${\cal O} (\beta)$ for the GUP-corrected AK system in the frame of non-relativistic quantum mechanics. 
In order to explore the AK inequality at the Planck scale, we need to introduce the initial state $\Psi_{in} (q_1, q_2, q_3: 0)$. Then, $\Psi (x_1, x_2, x_3: T)$
can be  directly derived by 
\begin{equation}
\label{concl-1}
\Psi (x_1, x_2, x_3: T) = \int dq_1 dq_2 dq_3 \bra{\bm{x}} \widehat{U}(T) \ket{\bm{q}} \Psi_{in} (q_1, q_2, q_3: 0).
\end{equation}
If we want to examine the effect of the initial entanglement of the probe system, we should choose $\Psi_{in} (q_1, q_2, q_3: 0) = \psi(q_1, q_2) \phi (q_3)$, where $\psi(q_1, q_2)$ is not separable. 
After deriving $\Psi (x_1, x_2, x_3: T)$ we can compute $\Delta_{\widehat{x}_1} (T = 1 / \kappa) \Delta_{\widehat{x}_2} (T = 1 / \kappa)$ and the corresponding AK inequality.
It is of interest to examine the effect of ML in the AK inequality. It may give some insight into the effect of the simultaneous measurement at early universe stage. 
We hope to explore this issue in the future.

%{\bf Acknowledgement}:
%On April 16, 2014 the ferry Sewol has sunk into the South Sea of Korea. Due to this disaster 304 people died and, 9 of them are still missing. We would like to dedicate this paper to all victims of this accident.
%This research was supported by the Basic Science Research Program through the National Research Foundation of Korea(NRF) funded by the Ministry of Education, Science and Technology(2011-0011971).
%This work was supported by the Kyungnam University Foundation Grant, 2017.

\newpage 

\begin{appendix}{\centerline{\bf Appendix A: Factorization of the time-evolution operator}}

\setcounter{equation}{0}
\renewcommand{\theequation}{A.\arabic{equation}}

Baker-Campbell-Hausdorff (BCH) formula is 
\begin{equation}
\label{app-1}
e^{\widehat{X}} e^{\widehat{Y}} = e^{\widehat{Z}}
\end{equation}
where $\widehat{Z} = \widehat{\Gamma}_0 + \widehat{\Gamma}_1 + \cdots$. First few $\widehat{\Gamma}_j$ are 
\begin{eqnarray}
\label{app-2}
&&\widehat{\Gamma}_0 = \widehat{X} + \widehat{Y}       \hspace{4.0cm} \widehat{\Gamma}_1 = \frac{1}{2} [ \widehat{X}, \widehat{Y} ]       \\     \nonumber
&& \widehat{\Gamma}_2 = \frac{1}{12} \left( [\widehat{X}, [\widehat{X}, \widehat{Y}]] + [\widehat{Y}, [\widehat{Y}, \widehat{X}]] \right)  \hspace{1.0cm} \widehat{\Gamma}_3 = \frac{1}{24} [\widehat{X}, [\widehat{Y}, [\widehat{Y}, \widehat{X}]]]        \\    \nonumber
&&  \widehat{\Gamma}_4 = -\frac{1}{720} \left([ \widehat{Y}, [\widehat{Y}, [\widehat{Y}, [\widehat{Y}, \widehat{X}]]]] + [ \widehat{X}, [\widehat{X}, [\widehat{X}, [\widehat{X}, \widehat{Y}]]]] \right)      \\    \nonumber
&&\hspace{1.0cm}  + \frac{1}{360} \left( [ \widehat{X}, [\widehat{Y}, [\widehat{Y}, [\widehat{Y}, \widehat{X}]]]] + [ \widehat{Y}, [\widehat{X}, [\widehat{X}, [\widehat{X}, \widehat{Y}]]]]  \right)     \\   \nonumber
&& \hspace{1.0cm}  + \frac{1}{120} \left(  [ \widehat{Y}, [\widehat{X}, [\widehat{Y}, [\widehat{X}, \widehat{Y}]]]] +  [ \widehat{X}, [\widehat{Y}, [\widehat{X}, [\widehat{Y}, \widehat{X}]]]]  \right).
\end{eqnarray}

Now, let us consider the factorization of the time-evolution operator $\widehat{U} (T) = e^{-i \widehat{H} T}$ up to ${\cal O}(\beta)$.
It is easy to show 
\begin{equation}
\label{app-3}
\widehat{U} (T) = e^{-\frac{i T}{2 m_1} \widehat{p}_1^2 - \frac{i \beta T}{m_1} \widehat{p}_1^4}  e^{-\frac{i T}{2 m_2} \widehat{p}_2^2 - \frac{i \beta T}{m_2} \widehat{p}_2^4} e^{\widehat{A} + \widehat{B}}
\end{equation}
where
\begin{eqnarray}
\label{app-4}
&&\widehat{A} = -i T \left[\frac{1}{2 m_3} \widehat{p}_3^2 + \kappa \widehat{p}_2 \widehat{p}_3 + \beta \left( \frac{1}{m_3} \widehat{p}_3^4 + \kappa \widehat{p}_2^3 \widehat{p}_3 + \kappa \widehat{p}_2 \widehat{p}_3^3 \right) \right]     \\   \nonumber
&&\widehat{B} = -i \kappa T \left( \widehat{q}_3 \widehat{p}_1 + \beta \widehat{q}_3 \widehat{p}_1^3 \right).
\end{eqnarray} 
Of course, the higher order terms of $\beta$ are disregarded in the exponent of Eq. (\ref{app-3}). Thus the main problem is to factorize $e^{\widehat{A} + \widehat{B}}$.

Making use of the first quantization rule $[\widehat{q}_i, \widehat{p}_j] = i$, the following non-vanishing commutators can be straightly derived:
\begin{eqnarray}
\label{app-5}
&& [\widehat{A}, \widehat{B}]= i \kappa T^2 \Bigg[\frac{1}{m_3} \widehat{p}_1 \widehat{p}_3 + \kappa \widehat{p}_1 \widehat{p}_2                            \\   \nonumber
&&   \hspace{3.0cm}   + \beta \left\{ \frac{1}{m_3} \left( \widehat{p}_1^3 \widehat{p}_3 + 4 \widehat{p}_1 \widehat{p}_3^3 \right) + \kappa \widehat{p}_1^3 \widehat{p}_2 + \kappa \widehat{p}_1 \widehat{p}_2^3 + 3 \kappa \widehat{p}_1 \widehat{p}_2 \widehat{p}_3^2  \right\}  \Bigg]    \\    \nonumber
&& [\widehat{B}, [\widehat{A}, \widehat{B}]] = i \kappa^2 T^3 \left[ \frac{1}{m_3} \widehat{p}_1^2 + 2 \beta \left\{ \frac{1}{m_3} \left( \widehat{p}_1^4 + 6 \widehat{p}_1^2 \widehat{p}_3^2 \right) + 3 \kappa \widehat{p}_1^2 \widehat{p}_2 \widehat{p}_3 \right\} \right]        \\    \nonumber
&& [\widehat{B}, [\widehat{B}, [\widehat{A}, \widehat{B}]]] = 6 i \beta \kappa^3 T^4 \left( \frac{4}{m_3} \widehat{p}_1^3 \widehat{p}_3 + \kappa \widehat{p}_1^3 \widehat{p}_2 \right)     \\     \nonumber
&& [\widehat{B},  [\widehat{B}, [\widehat{B}, [\widehat{A}, \widehat{B}]]]] = \frac{24 i \beta}{m_3} \kappa^4 T^5 \widehat{p}_1^4.
\end{eqnarray}
Of course, the higher terms of $\beta$ are ignored in Eq. (\ref{app-5}). Since other commutators are ${\cal O}(\beta^2)$, we will set to zero in the following computation. 
Applying the BCH formula (\ref{app-1}) one can show straightforwardly
\begin{eqnarray}
\label{app-6}
e^{\frac{1}{2} \widehat{A}} e^{\widehat{B}} e^{\frac{1}{2} \widehat{A}} &=& \exp \left[ \widehat{A} + \widehat{B} - \frac{1}{12} [\widehat{B}, [\widehat{A}, \widehat{B}]] + \frac{1}{720}  [\widehat{B},  [\widehat{B}, [\widehat{B}, [\widehat{A}, \widehat{B}]]]]  \right]     \\    \nonumber
&=& e^{\widehat{A} + \widehat{B} - \frac{1}{12} [\widehat{B}, [\widehat{A}, \widehat{B}]]} e^{\frac{1}{720}  [\widehat{B},  [\widehat{B}, [\widehat{B}, [\widehat{A}, \widehat{B}]]]]}.
\end{eqnarray}
The last equality is due to the fact that $[\widehat{B},  [\widehat{B}, [\widehat{B}, [\widehat{A}, \widehat{B}]]]]$ commutes with other quantities because it does not contain the factor $\widehat{p}_3$.
Rearranging Eq. (\ref{app-6}) one can derive 
\begin{equation}
\label{app-7}
e^{\widehat{F} + \widehat{G}} = e^{\frac{i \kappa^2 T^3}{12} \left(\frac{1}{m_3} \widehat{p}_1^2 + \frac{2 \beta}{m_3} \widehat{p}_1^4 \right)} e^{-\frac{1}{720}  [\widehat{B},  [\widehat{B}, [\widehat{B}, [\widehat{A}, \widehat{B}]]]]} 
e^{\frac{1}{2} \widehat{A}} e^{\widehat{B}} e^{\frac{1}{2} \widehat{A}}
\end{equation}
where
\begin{equation}
\label{app-8}
\widehat{F} = \widehat{A} + \widehat{B}         \hspace{1.0cm}  \widehat{G} = - \frac{i \beta \kappa^2 T^3}{2} \left( \frac{2}{m_3} \widehat{p}_1^2 \widehat{p}_3^2 + \kappa \widehat{p}_1^2 \widehat{p}_2 \widehat{p}_3 \right).
\end{equation}
One can show the followings again:
\begin{eqnarray}
\label{app-9}
&& [\widehat{F}, \widehat{G} ] = - \frac{i \beta \kappa^3 T^4}{2} \left( \frac{4}{m_3} \widehat{p}_1^3 \widehat{p}_3 + \kappa \widehat{p}_1^3 \widehat{p}_2 \right)                  \\    \nonumber
&&[\widehat{F}, [\widehat{F}, \widehat{G}]] = - \frac{2 i \beta \kappa^4 T^5}{m_3} \widehat{p}_1^4.
\end{eqnarray}
Applying the BCH formula again, one can show 
\begin{equation}
\label{app-10}
 e^{\widehat{F} + \widehat{G}} = e^{\frac{1}{2} \widehat{G}} e^{\widehat{F}} e^{\frac{1}{2} \widehat{G}} e^{-\frac{1}{12} [\widehat{F}, [\widehat{F}, \widehat{G}]]}
 \end{equation}
Inserting Eq. (\ref{app-10}) into Eq. (\ref{app-7}), one can compute $e^{\widehat{F}}$ as following;
\begin{equation}
\label{app-11}
e^{\widehat{F}} = e^{\widehat{A} + \widehat{B}}
= e^{-\frac{1}{2} \widehat{G}} e^{\frac{i \kappa^2 T^3}{12} \left(\frac{1}{m_3} \widehat{p}_1^2 + \frac{2 \beta}{m_3} \widehat{p}_1^4 \right)} e^{-\frac{1}{720}  [\widehat{B},  [\widehat{B}, [\widehat{B}, [\widehat{A}, \widehat{B}]]]]} 
e^{\frac{1}{2} \widehat{A}} e^{\widehat{B}} e^{\frac{1}{2} \widehat{A}} e^{\frac{1}{12} [\widehat{F}, [\widehat{F}, \widehat{G}]]} e^{-\frac{1}{2} \widehat{G}}.
\end{equation}
Inserting Eq. (\ref{app-11}) into Eq. (\ref{app-3}) one can successfully factorize the time-evolution operator as following form:
\begin{equation}
\label{app-12}
\widehat{U} (T) = e^{\Delta x_1 \widehat{p}_1^2 + \beta \Delta y_1 \widehat{p}_1^4} e^{-\frac{i T}{2 m_2} \widehat{p}_2^2 - \frac{i \beta T}{m_2} \widehat{p}_2^4} \widehat{V} (T) e^{-i \kappa T \widehat{q}_3 (\widehat{p}_1 + \beta \widehat{p}_1^3)} 
e^{-\frac{i \beta \kappa^4 T^5}{6 m_3} \widehat{p}_1^4} \widehat{V} (T)
\end{equation}
where 
\begin{eqnarray}
\label{app-13}
&&\Delta x_1 = -\frac{i T}{2 m_1} + \frac{i \kappa^2 T^3}{12 m_3}   \hspace{1.0cm} \Delta y_1 = - \frac{i T}{m_1} + \frac{i \kappa^2 T^3}{6 m_3} - \frac{i \kappa^4 T^5}{30 m_3}       \\     \nonumber
&&\widehat{V} (T) = e^{-\frac{i T}{4 m_3} \widehat{p}_3^2 - \frac{i \beta T}{2 m_3} \widehat{p}_3^4} e^{-\frac{i \kappa T}{2} \left[ \widehat{p}_2 \widehat{p}_3 + \beta (\widehat{p}_2^3 \widehat{p}_3 + \widehat{p}_2 \widehat{p}_3^3 ) \right]}
e^{\frac{i \beta \kappa^2 T^3}{2 m_3} \widehat{p}_1^2 \widehat{p}_3^2} e^{\frac{i \beta \kappa^3 T^3}{4} \widehat{p}_1^2 \widehat{p}_2 \widehat{p}_3}.
\end{eqnarray}

\end{appendix}

\newpage 

\begin{appendix}{\centerline{\bf Appendix B: Coefficients $f_{ijk}$}}

\setcounter{equation}{0}
\renewcommand{\theequation}{B.\arabic{equation}}
In this appendix we summarize the coefficients $f_{ijk}$ arising in Eq. (\ref{feynman-13}). 
Several coefficients are as following;
\begin{eqnarray}
\label{app-b-1}
&&f_{400} = -432 i m_1^3 m_3^3 b^3 \left[ 20 a(T) + 3 m_1 \kappa^4 T^4 \right]   \hspace{.5cm} f_{040} = 5 i m_2^3 a^4 (T)         \\    \nonumber
&&f_{301} = -2592 i m_1^3 m_3^3 \kappa T b^3 \left[5 a(T) + m_1 \kappa^4 T^4 \right]  \hspace{.5cm} f_{031} = -5 i m_2^3 m_3 \kappa a^4(T) (3 + m_3^2 \kappa^2 )   \\   \nonumber
&&f_{220} = -180 i m_1^2 m_2^2 m_3^3 \kappa^4 T^2 b a^2(T)  \hspace{.5cm} f_{112} = -180 i m_1^2 m_2 m_3^3 \kappa^4 T^3 (b - 2) b a^2 (T)             \\   \nonumber
&& f_{121} = -180 i m_1^2 m_2^2 m_3^3 \kappa^5 T^3 b a^2 (T)  \hspace{.5cm} f_{211} = -180 i m_1^2 m_2 m_3^3 \kappa^3 T^2 (b - 2) b a^2 (T)           \\    \nonumber
&&f_{300} = 5184 i m_1^3 m_3^3 Q_3 \kappa T b^3 \left[ 5 a(T) + m_1 \kappa^4 T^4 \right] \hspace{.5cm} f_{120} = 360 i m_1^2 m_2^2 m_3^3 Q_3 \kappa^5 T^3 b a^2 (T)    \\    \nonumber
&&f_{201} = 2592 i m_1^3 m_3^3 Q_3 \kappa^2 T^2 b^3 \left[ 10 a(T) + 3 m_1 \kappa^4 T^4 \right] \hspace{.5cm} f_{021} = 180 i m_1^2 m_2^2 m_3^3 Q_3 \kappa^6 T^4 b a^2 (T)           \\    \nonumber
&&f_{012} = 180 i m_1^2 m_2 m_3^3 Q_3 \kappa^5 T^4 (b - 2) b a^2 (T)   \hspace{.5cm} f_{111} = 360 i m_1^2 m_2 m_3^3 Q_3 \kappa^4 T^3 (b - 2) b a^2 (T).
\end{eqnarray}
Other nonzero coefficients have long expressions. Of course they can be read by performing the integration of Eq. (\ref{feynman-11}). 

Now, we define 
\begin{equation}
\label{app-b-2}
h_{ijk} = \frac{f_{ijk}}{5 T^3 b^3 a^4 (T)} \Bigg|_{\kappa = 0}.
\end{equation}
Then, the nonzero $h_{ijk}$ are
\begin{eqnarray}
\label{app-b-3}
&&  h_{400} = -i \frac{m_1^3 }{T^3}   \hspace{.5cm}  h_{040} =  -i \frac{m_2^3 }{T^3}    \hspace{.5cm}  h_{004} =  -i \frac{m_3^3 }{T^3}               \\   \nonumber
&& h_{200} = - \frac{6 m_1^2}{T^2}   \hspace{.5cm} h_{020} = - \frac{6 m_2^2}{T^2}   \hspace{.5cm} h_{002} = - \frac{6 m_3^2}{T^2}    \\   \nonumber       
&&   \hspace{1.5cm} h_{000} = \frac{3 i (m_1 + m_2 + m_3)}{T}.      
\end{eqnarray}

\end{appendix}

\end{document}